\documentclass[
reprint, twocolumn, secnumarabic%
,amssymb,
nobibnotes, nofootinbib, aps, prl, showpacs,showkeys]{revtex4}
\usepackage{docs}%
\usepackage{bm}%
\usepackage{graphicx}
\usepackage{rotating}
\usepackage{hyperref}
\usepackage{url}
\usepackage{srcltx}
\expandafter \ifx \csname package@font\endcsname\relax\else
\expandafter \expandafter \expandafter
\usepackage
\expandafter \expandafter
\expandafter{\csname package@font\endcsname}%
\fi



\usepackage{enumerate}

\begin{document}

\title{Relevance of Two Boson Exchange Effect in Quasi-Elastic Charged Current Neutrino-Nucleon Interaction}

\author{Krzysztof M. Graczyk}

\affiliation{Institute of
Theoretical Physics, University of Wroc\l aw, pl. M. Borna 9,
50-204, Wroc\l aw, Poland}

\begin{abstract}
Two boson exchange (TBE) correction to the cross section for the
quasi-elastic charged current $\nu n$ and $\overline{\nu} p$
scattering is evaluated. The TBE is given by $W\gamma$ box
diagrams. The calculations are performed for 1 GeV neutrinos. The
averaged TBE correction is of the order of $2\div4\%$ (with
respect to Born contribution) in the case of $\nu_e$ and
$\overline{\nu}_e$ and  $1\div2\%$ in the case of $\nu_\mu$ and
$\overline{\nu}_\mu$. The impact of the TBE effect on the
systematic discrepancy between the $\nu_e$ and $\nu_\mu$ cross
sections  is discussed.

\end{abstract}

\maketitle

\section{Introduction}

One of the goals of the particle physics is to understand the
fundamental properties of neutrinos. In the experimental physics
an effort has been made to measure  the $\theta_{13}$ parameter
and then, in the near future, the CP violation phase. It can be
done by analyzing the  $\nu_\mu \to \nu_e$, $\overline{\nu}_\mu
\to \overline{\nu}_e$, $\nu_e \to \nu_\mu$ and $\overline{\nu}_e
\to \overline{\nu}_\mu$ oscillation processes.

Estimate of the systematic differences between the cross sections
for $\nu_e$ and $\nu_\mu$ interactions is of importance  in the
reconstructing $\nu_e$ signal \cite{Day:2012gb}. Electrons,
because $m_e \ll m_\mu$, have tendency to radiate more than muons.
Therefore radiative corrections (RCs) are the potential source of
the discrepancy between cross sections for the charged current
electron and muon neutrinos interactions. Moreover the RCs are not
included in any Monte Carlo generator used to analyze the neutrino
interaction data \cite{Day:2012gb}.

In the long base-line experiments like T2K \cite{Abe:2011sj} or
NO$\nu$A \cite{Nowak:2012zz} the charged current quasi-elastic
(CCQE) neutrino-nucleon scattering is the dominant observed
process. The typical averaged neutrino energy in those experiments
is about 0.7 GeV and 2 GeV, while the typical four momentum
transfer $Q^2<4$ GeV$^2$. In this paper we  concentrate our
attention  on the study of the impact of  the two boson exchange
(TBE) effect on the CCQE cross sections in this kinematical
domain.
\begin{figure}
\centering{
\includegraphics[scale=0.5]{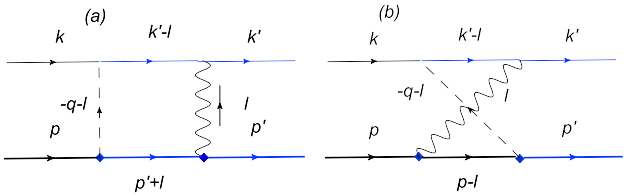}
\caption{ Two boson exchange $W\gamma$ box diagrams.
\label{Fig_Wplusg}} }
\end{figure}

 Recently the TBE effect has been extensively investigated in the elastic
 $ep$ scattering. The renewal interest of this topic was induced by
 observing the discrepancy between  the measurements of the
 form factor ratio $G_E^p/G_M^p$ ($G_{E,M}^p$ is the electric,
 magnetic proton form factor) obtained within two different
 experimental techniques.  The first method is based on the
 Rosenbluth separation, the other is based on the  so-called
 polarization transfer (PT) measurements (for review see
 Ref. \cite{Arrington:2011dn}).

 In the electron scattering the leading contribution to the TBE
 effect is given by interference of the Born amplitude with the
 $\gamma\gamma$ box diagrams describing an exchange of two virtual
 photons between electron and the target. It is called two photon
 exchange (TPE) correction. In the classical treatment of RCs to $ep$
  scattering \cite{MT_radiative} the TPE correction was  estimated
  in an approximation, in which the hard photon contribution,
  induced by the internal proton structure, was neglected.
  It turned out that applying in the Rosenbluth analysis the new, more precise predictions of
  the TPE effect, allowed to partially resolve  the problem of
  discrepancy between the form factors
  \cite{Blunden:2003sp,tpe_first_explenation}.

 In the case of neutrino interactions the RCs have been extensively
 studied for $\beta$ and $\mu $ decays  but also for the CCQE
 neutrino-deuteron interaction near the threshold
 region \cite{radiative_neutrino,Fukugita:2004cq}.
 The RCs were also estimated for the  deep inelastic (DIS) $\nu$-nucleon scattering \cite{Dis,De
Rujula:1979jj}.  The total corrections are of the order of several
percent in the threshold region. The size of the RCs in the DIS
depends on kinematics, measurement technique
 (e.g. the way of reconstructing  $Q^2$),
 detector properties, kinematical cuts etc.

 Recently some discussion of the impact of the RCs
 on the CCQE cross sections in 1 GeV kinematical domain was presented in
 Ref. \cite{Day:2012gb,Bodek:2007wb}, where the lepton leg correction formula, estimated
 in the leading log approximation \cite{De Rujula:1979jj} was adopted. It describes the
 soft and hard photon emission by the charged lepton leg ($\nu p$, DIS) and it does not include the TBE contribution. Obviously this approximation is
based on the diagrams,  which do not form the gauge invariant set.

 The aim of this paper is to discuss the contribution given by $W\gamma$ box diagrams induced by the internal nucleon structure.
 Certainly  the predictions can be model dependent. The simplest way is to consider the box diagrams as drown in Fig. \ref{Fig_Wplusg} and then to proceed as it follows:
 (i) assume that the hadronic
 intermediate state is given by the off-shell nucleon;  (ii) the off-shell electroweak hadronic vertices are modelled by the on-shell nucleon form factors.
 This kind of the approximation was successfully applied for predicting the hard photon contribution from
  $\gamma\gamma$ \cite{Blunden:2005ew} and $\gamma Z^0$ \cite{Zhou:2007hr,Zhou:2009nf,Tjon:2009hf} box diagrams.
A part of the inner radiative correction correction to the $\beta$
decay was also estimated by considering the nucleon form factors
\cite{Marciano:1985pd,Fukugita:2004cq}.

This approach seems to lead to the reasonable predictions of the
TPE effect for $ep$ scattering in the low and the intermediate
$Q^2$ range \cite{Arrington:2007ux}. Hence in the same kinematical
domain as it is considered in the case of the 1 GeV neutrino
interactions. In one of our previous papers \cite{Graczyk:2013pca}
we made an effort to predict the TPE correction for $\gamma\gamma$
box contribution in the elastic $ep$ scattering, considering
nucleon and $\Delta(1232)$ resonance as the intermediate hadronic
states. We showed the satisfactory agreement between the
theoretical results and phenomenological fits obtained from the
Bayesian analysis of the $ep$ scattering data
\cite{Graczyk:2011kh}. In this paper we apply the same methodology
to compute
 $W\gamma$ contribution in CCQE reactions.

\section{Formalism}

Let us consider quasi-elastic  charged current neutrino-nucleon
scattering $\nu_l (k) + n(p)\to  l^-(k') + p(p')$, where $l =e$,
$\mu$. We define the kinematical variables $q^\mu = k^\mu -
{k'}^\mu= (q_0,\mathbf{q})$ (four momentum transfer), $Q^2 =
-q^2$. Mandelstam variables read $s=(k+p)^2$ and $t=(k-k')^2=q^2$,
while $E$ and $E'$ denotes the neutrino and lepton energy.

A Born amplitude for the CCQE $\nu n$ reaction reads,
\begin{equation}
i\mathcal{M}_{Born} =
i\frac{g^2 \cos\theta_C  }{8 (t - M_W^2)} j_\mu h^\mu,
\end{equation}
$\theta_C=13.04^\circ$  is Cabbibo angle, $g= e/\sin\theta_W$ is the weak coupling constant, $\sin^2\theta_W=0.2312$, $e^2=  4\pi \alpha$, $\alpha=1/137$, $M_W=80.3$ GeV is the boson $W^\pm$ mass. We mostly follow the convention of Cheng and Lee textbook \cite{Cheng_book}.

The leptonic one-body current has a simple form,
\begin{equation}
j_\mu = \overline{u}(k')\gamma_\mu(1-\gamma_5)u(k),
\end{equation}
while the hadronic one-body current reads,
\begin{equation}
h^\mu(q) = \overline{u}(p') \Gamma^\mu_{cc}u(p).
\end{equation}
The electorweak nucleon vertex is a function of four form factors,
\begin{equation}
\Gamma^\mu_{cc}(q\equiv p'-p) = \Gamma^\mu_V(q)  -
\gamma_\mu\gamma_5 F_A(q) -\frac{q^\mu \gamma_5}{2M} {F_P(q)},
\end{equation}
where, $\Gamma_V^\mu = \Gamma^\mu_p(q) -\Gamma^\mu_n(q)$,
$\Gamma^\mu_{p(n)}(q)$ is the proton (neutron) electromagnetic
vertex defined below, while $F_A$ and $F_P$ are the nucleon and
pseudoscalar axial form factors respectively. We assume that
$F_A(q) =g_A /(1 +Q^2/M_A^2)^2$, where $g_A = 1.267$, $M_A$ is an
axial mass, and as a default value we take $M_A=1$ GeV.  The
pseudoscalar axial form factor, $F_P$ has a commonly used form:
$F_P(q) =  4 M^2 F_A(q) /(m_\pi^2- q^2 )$; $m_\pi$ is the pion
mass, $M=(M_p+M_n)/2$, $M_{p,(n)}$ is the proton (neutron) mass.

The proton (neutron) electromagnetic vertex reads,
\begin{equation}
\Gamma_{p,n}^\mu(q) = \gamma^\mu F_1^{p,n}(q) + \frac{i\sigma^{\mu\nu} q_\nu}{2M_{p,n}} F_2^{p,n}(q),
 \end{equation}
 where $F_{1,2}^{p(n)}$ is proton (neutron) form factor.
\begin{figure}
\centering{
\includegraphics[width=0.35\textwidth]{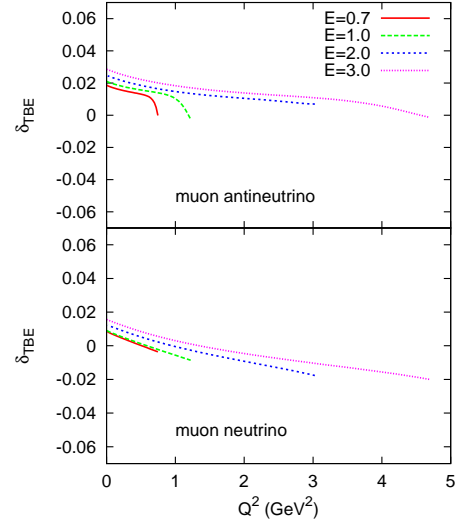}
\caption{$\delta_{TBE}$ for $\nu_\mu n$ and $\overline{\nu}_\mu p$
CCQE reactions. \label{Fig_delta_TBE_muon}}}
\end{figure}

 It is convenient to express the nucleon electromagnetic form factors by  the electric and magnetic proton (neutron) form factors,
$F_1^{p,n}  =\left( G_E^{p,n} +\tau_{p,n} G_M^{p,n}\right)/(1 +
\tau_{p,n})$, $F_2^{p,n} = \left( G_M^{p,n}-  G_E^{p,n}\right)/(1
+ \tau_{p,n})$, where $\tau_{p,n} = - {q^2}/{4M^2_{p,n}}$. In our
calculations we consider a dipole parametrization of the electric
and magnetic form factors, namely, $G_E^p(Q^2) =
G_M^{p,n}(Q^2)/\mu_{p,n} = \Lambda^4/(Q^2+\Lambda^2)^2$, where
$\Lambda=0.84$ GeV is the cut off parameter. The electric neutron
form factor is assumed to be zero ($G_E^n(Q^2)=0$).

\begin{figure}
\centering{
\includegraphics[width=0.35\textwidth]{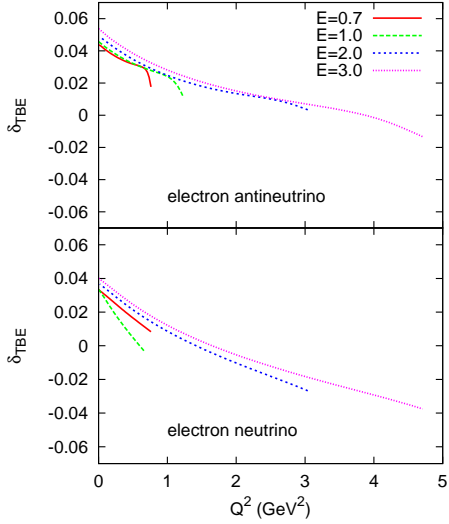}
\caption{Similarly as in Fig. \ref{Fig_delta_TBE_muon} but for
electron neutrinos. \label{Fig_delta_TBE_electron}}}
\end{figure}

In order to calculate  the TBE correction we consider the exchange
between the lepton and the nucleon target the virtual $W^+$ boson
and the photon $\gamma$. For the hadronic intermediate state we
take the off-shell nucleon (N). We expect that the resonance
hadronic contribution, similarly, as in the case of $ep$
scattering will be very small
\cite{Kondratyuk:2005kk,Graczyk:2013pca}, in particular, in the
low $Q^2$ range, which is the most relevant domain for neutrino
reactions.

Two box $W\gamma$ diagrams contribute to the TBE amplitude, see
Fig. \ref{Fig_Wplusg}, $i \Box_{ W^+ \gamma}^\| = - {\cos\theta_C
e^2  g^2} I_{ W^+ \gamma}^\| /{8} $ and $i \Box_{
W^+\gamma}^\times=  - {\cos\theta_C e^2  g^2} I_{
W^+\gamma}^\times/8$, where
\begin{eqnarray}
I_{ W^+ \gamma}^\|
 &=& \int \frac{d^4 l}{(2\pi)^4}
 \frac{  l^{\mu\nu} h_{\mu\nu}^\parallel}{D(p',M_p)} \\
I_{  W^+\gamma}^\times
 &=& \int \frac{d^4 l}{(2\pi)^4}
 \frac{ l^{\mu\nu} h_{\mu\nu}^\times}{D(-p,M_n)}\\
 \nonumber\\
 l^{\mu\nu}
 &=&  \overline{u}(k') \gamma^\mu   (\hat{k'} - \hat{l} + m)  \gamma^\nu(1-\gamma_5)  u(k)\nonumber\\
 h_{\mu\nu}^\parallel
 &=&\overline{u}(p')  \Gamma_\mu^{p}(-l)  (\hat{p'} + \hat{l} + M_p)\Gamma_\nu^{cc} (q+l)   u(p)\nonumber \\
 h_{\mu\nu}^\times
 &=& \overline{u}(p') \Gamma_\nu^{cc}(q+l) (\hat{p} - \hat{l} + M_n)  \Gamma_\mu^{n}(-l)     u(p) \nonumber
\end{eqnarray}
where $D(x,M_x) = [(q+l)^2 - M_W^2+ i\epsilon][ l^2 + i\epsilon] [(k'-l)^2 - m^2 + i \epsilon][(x+l)^2 - M^2_x+ i \epsilon]$, $m$ denotes the lepton mass.

The TBE correction to spin averaged cross section is the
interference of the Born and TBE box diagrams, and it reads,
    \begin{eqnarray}
    \Delta_{TBE} &=&   \mathrm{Re} \sum_{spin}   \mathcal{M}_{Born}^* \left(  \Box_{ W^+ \gamma}^\|  +  \Box_{ W^+ \gamma  }^\times\right)
    \\
    &=& \frac{g^4 e^2 \cos^2\theta_C }{16 (M_W^2-t)}
    \underbrace{\mathrm{Im}\sum_{spin}  (j_\alpha h^\alpha)^*
    \left(I_{W^+ \gamma}^\| + I_{ W^+
    \gamma}^\times\right)}_{\diamondsuit} \nonumber
    \end{eqnarray}
In practice $\diamondsuit  \sim \int d^4 l \; N/D$, where $N$ is a
a polynomial function of: $l^2$, $l\cdot p'$, $l\cdot k'$ and
$l\cdot q$ scalar products, given by an appropriate sum of traces,
computed with FeyCalc package \cite{FeynCalc}, while $D$ denotes
denominator, which because of the form factors can be of the order
of $l^8$.

To compute the TBE contribution the integral $\diamondsuit$ was
expressed as a sum of scalar loop integrals
\cite{Veltman_Passarino_Integrals}. But to perform this
decomposition the appropriate reduction of $N$ with $D$ had to be
done.  Eventually, the numerical values of the TBE correction were
evaluated applying LoopTool  library (C++) \cite{looptool}.

Both box amplitudes, because of the form factors, are ultraviolet
finite. The amplitude $i \Box_{W^+\gamma}^\parallel$ is infrared
(IR) divergent, while $i \Box_{W^+\gamma}^\times$ amplitude is IR
finite. It is easy to show, in the limit $l \to 0$ $i\Box_{ W^+
\gamma  }^\times$ behaves as
\begin{eqnarray}
& \sim  & \mu_n \int  \frac{d l l^3}{l^2 \; k'\cdot l \; p'\cdot l  }  \left(\frac{\gamma^\mu  l^2}{l^2 -4M^2_{n} }  +  \sigma^{\mu\nu}l_\nu\right).
\nonumber
\end{eqnarray}
\begin{figure}
\centering{
\includegraphics[width=0.35\textwidth]{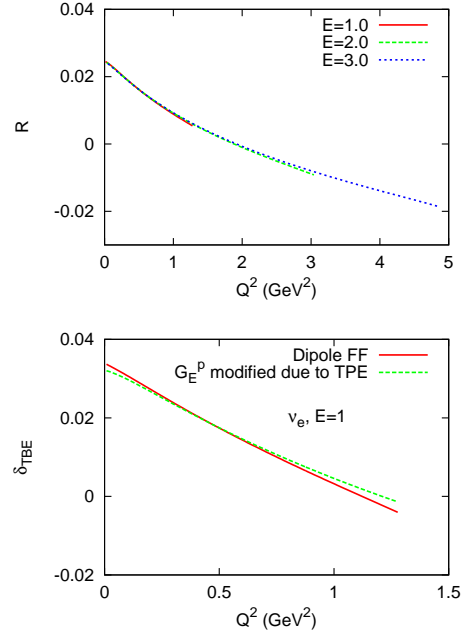}
\caption{Top panel: ratio  (\ref{R_ratio}) for $\nu_e$ of energy
of 1 GeV. Bottom panel: $\delta_{TBE}$ computed for the dipole
electromagnetic form factors and for the form factors modified due
to the TPE effect (Eq. \ref{GEp_TPE}). \label{Fig_ratio}}}
\end{figure}

\begin{figure}
\centering{
\includegraphics[width=0.35\textwidth]{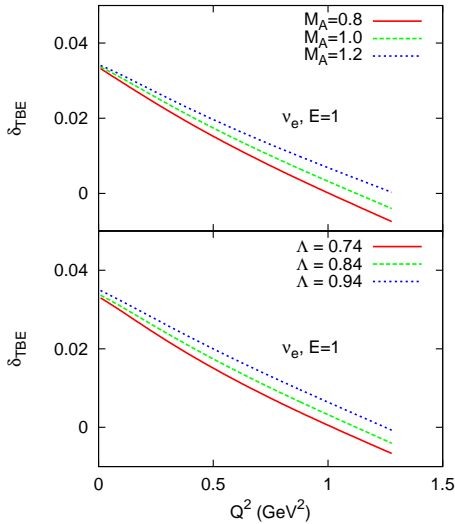}
\caption{$\delta_{TBE}$ dependence on the hadronic model
parameters, the axial mass $M_A$ (top panel) and the cut-off
vector form factor parameter $\Lambda$ (bottom
panel).\label{Fig_dependence}}}
\end{figure}

We extract the hard photon contribution, which is related to the
internal structure of the nucleon, by subtracting the IR divergent
contribution (soft photon contribution).
\begin{equation}
\label{delta_TBE}
\delta_{TBE} = \frac{\Delta_{TBE} -\Delta_{TBE}(\Box_{ W^+ \gamma}^\|,soft)}{\frac{1}{2}\sum_{spin}|\mathcal{M}_{Born}|^2}
\end{equation}
Similar regularization procedure was discussed in Refs.
\cite{Blunden:2003sp,Zhou:2007hr}.

The soft contribution reads,
\begin{equation}
\Delta_{TBE}^{\nu n }(\Box_{ W^+ \gamma}^\|,soft) =
   -4 k'\cdot p' e^2 |\mathcal{M}_{Born}|^2  \mathrm{Im}C_0,
\end{equation}
$C_0 \equiv C_0(m^2,M_p^2,s,m^2,0,M_p^2)$ is three-point scalar one-loop integral \cite{Veltman_Passarino_Integrals},
 \begin{eqnarray}
C_0&=&
\int \frac{d^4 l}{(2\pi)^4}
 \frac{ 1}{ l^2 [(k'-l)^2 - m^2 ][(p'+l)^2 - M^2_p ]}
 \end{eqnarray}
In order to deal with the IR divergencies, the photon mass $\mu$
is introduced $1/l^2 \to 1/(l^2 - \mu^2)$. Notice that the IR
divergency coming from $W\gamma$ box diagram is cancelled, in the
total calculus of the RCs, by the soft photon Bremsstrahlung
inelastic contribution.

The calculations for the antineutrino scattering are
straightforward. The leptonic current in this case reads,
$j_\mu^{cc} = \overline{u}(k') \gamma_\mu (1+\gamma_5)u(k)$ ,
while $l^{\mu\nu}= -\overline{u}(k')\gamma^\nu  (\hat{k'} -
\hat{l} + m)  \gamma^\mu (1+\gamma_5) u(k)$. One should also
interchange the electromagnetic vertices in the diagrams
$\Gamma_p^\mu \leftrightarrows \Gamma_n^\mu$. Notice that the
direct and exchange diagrams are interchanged.

\section{Discussion}

As mentioned in the introduction the  radiative corrections are
the potential source of systematical difference between the
electron and muon neutrino cross sections. In fact, the TBE effect
for $\nu_e$ is about two times larger than in the case of
$\nu_\mu$ (see Figs. \ref{Fig_delta_TBE_muon} and
\ref{Fig_delta_TBE_electron}). At low $Q^2$ the TPE correction is
positive, but when $Q^2$ grows it changes a sign.

In Fig. \ref{Fig_ratio} (top panel) we plot the quantity
\begin{eqnarray}
\mathcal{R}(Q^2) &=&
\frac{d\sigma_{Born}^{\nu_e}+d\sigma_{TBE}^{\nu_e}}{d\sigma_{Born}^{\nu_\mu}+d\sigma_{TBE}^{\nu_\mu}}
\left(\frac{d\sigma_{Born}^{\nu_e}}{d\sigma_{Born}^{\nu_\mu}}\right)^{-1} -1 \nonumber \\
 & = &
\frac{1+\delta_{TBE}(\nu_e)}{1+\delta_{TBE}(\nu_\mu)} -1 \approx
\delta_{TBE}(\nu_e)-\delta_{TBE}(\nu_\mu),\nonumber \\
\label{R_ratio}
\end{eqnarray}
which measures the relative difference between the TBE effect for
$\nu_e$ and $\nu_\mu$.  It  turned out to be a linear function in
$Q^2$, which takes the largest values at low $Q^2$.

Similarly as in the case of the TPE effect, in elastic $ep$
scattering \cite{Blunden:2003sp}, the TBE correction weakly
depends on the model parameters. It is  shown  in Fig.
\ref{Fig_dependence}, where we plot the TBE contribution computed
for several values of $M_A$ and $\Lambda$.  Moreover,  taking into
consideration more realistic electric proton form factor, modified
due to  TPE effect \cite{Alberico:2008sz},
\begin{equation}
\label{GEp_TPE} G_E^p(Q^2) = (-0.130\, Q^2 + 1.0022)
G_M^p(Q^2)/\mu_p
\end{equation}
has also a minor impact on the TBE effect, see Fig.
\ref{Fig_ratio} (bottom panel).

\begin{figure}
\centering{
\includegraphics[width=0.4\textwidth]{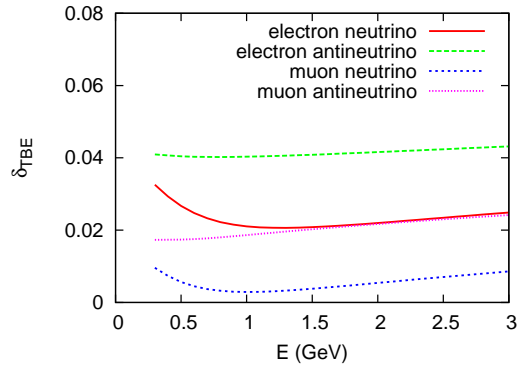}
\caption{Fraction of the TBE correction to the total cross
sections. \label{Fig_total}}}
\end{figure}

On average, the TBE correction is of the order of $2$ and $4$
percent for $\nu_e$ and $\overline{\nu}_e$ respectively. In the
case of the $\nu_\mu$ the TBE effect is negligible, while for
$\overline{\nu}_\mu$ it increases the total cross section by about
$2\%$, see Fig. \ref{Fig_total}. Notice that the axial mass $M_A$,
is usually fit to the total $\nu_\mu$ CCQE cross section data.
Therefore in order to re-construct, due to the TBE effect, the
''true'' $\nu_e$ CCQE cross section one should decrease $M_A$ by
about $2\%$.

More detailed investigation  of the impact of the TBE effect on
the cross sections requires considering the rest of the radiative
corrections. Taking into account the full set of diagrams required
by the Standard Model  goes beyond the scope of this paper.
However, we consider  the QED-like corrections, namely, the soft
Bremsstrahlung photon emission by the charged lepton and the
proton and the propagator corrections (for electron and proton).

The soft photon Bremsstrahlung emission describes the inelastic
process $\nu + n \to l^- + p + \gamma$, which   is
indistinguishable with the CCQE reaction as long as the photon has
an energy smaller than detector acceptance $\Delta E$. Hence it
contributes to the CCQE cross section. Notice that $\Delta E =
E'(quasi-elastic) - E'(inelastic) $. In our computations we
consider $\Delta E = \omega E'$ where $\omega \ll 1$.

The Bremsstrahlung corrections are  computed in the similar way as
in Ref. \cite{Maximon:2000hm}.  The soft photon emission cross
section reads,
\begin{eqnarray}
d \sigma_{Soft. Brem.} & \approx & d \sigma_{Born} \delta_{Soft. Brem.} \nonumber \\
\delta_{Soft. Brem.}  &=& \frac{\alpha}{4\pi^2} \left[2 {p'}\cdot
{k'} L_{p'k'} -
m^2 L_{k'k'} - M^2_pL_{p'p'} \right], \nonumber \\
\end{eqnarray}
where
\begin{equation}
\label{Brem_integral} L_{xy}= {\int_{|\mathbf{l}|<\Delta
\mathcal{E}}}\frac{d^3 l}{|\mathbf{l}|}\frac{1}{(x\cdot l)(y\cdot
l)} ,
\end{equation}
 $\Delta \mathcal{E}$ is the maximal photon energy in the frame with $\mathbf{l}+\mathbf{p'}=0$\footnote{In
the case of the electron the relation between $\Delta E$ and
$\Delta \mathcal{E}$ reads $\Delta \mathcal{E} =
(1+(E/M_p)(1-\cos\theta)) \Delta E$. When the charged lepton is
given by the muon the relation becomes a little complicated,
because the presence of the muon mass.}
 \cite{Veltman_Passarino_Integrals}. Certainly above integral is
 divergent when $l\to 0$.

The charged lepton  and the proton propagator corrections are easy
to derive and they read,
\begin{eqnarray}
\delta_{Self.} &=&\nonumber
\\
&
&\!\!\!\!\!\!\!\!\!\!\!\!\!\!\!\!\!\!\!\!-\frac{\alpha}{\pi}\left\{\frac{9}{4}
+ \ln\frac{\mu}{m} + \ln\frac{\mu}{M_p} +
\frac{1}{2}\left(\ln\frac{\Lambda_{UV}}{m}+\ln\frac{\Lambda_{UV}}{M_p}\right)\right\}.\nonumber
\\
\end{eqnarray}
The total correction: $\delta=\delta_{TBE}+\delta_{Soft. Brem.} +
\delta_{Self.}$  is IR finite, however, the propagator correction
depends on the UV cut-off parameter $\Lambda_{UV}$. We set
$\Lambda_{UV}=M_p$.

\begin{figure}
\centering{
\includegraphics[width=0.4\textwidth]{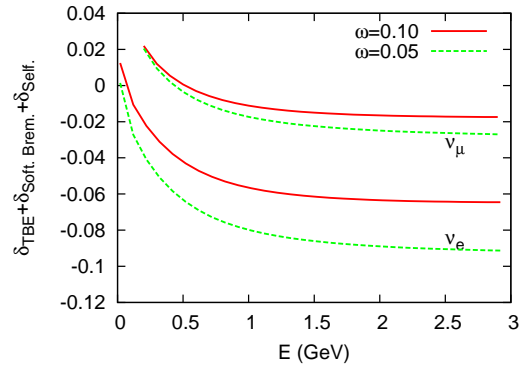}
\caption{$\delta_{TBE} + \delta_{Soft.Brem.}+\delta_{Self.}$ the
QED-like corrections. Only the soft photon Bremsstrahlung taken
into account. Two upper lines correspond to the results for
$\nu_\mu$, while two lower lines denote the computation done for
$\nu_e$. \label{Fig_total_brem}} }
\end{figure}

Fig. \ref{Fig_total_brem} shows $\delta$  for $\omega=0.05$ and
$\omega=0.1$. In the case of $\nu_e$ the correction is relatively
large and it is of the order of $-9\%$, while in the case of
$\mu_\mu$ it is around $-2\%$. The dominant (negative)
contribution comes from the soft Bremsstrahlung part. It leads to
the reduction of $\sigma(\nu_e)/\sigma(\nu_\mu)$ ratio by about
$-6\%$.

Above estimate gives the lower bound for the RCs. Adding the hard
photon Bremsstrahlung contribution (it is the typical way of
presenting the RCs in $\nu N$ scattering) reduces the $\delta$ and
it makes the RCs positive.

For complete calculus of the total RCs one should consider full
set of the diagrams required by the gauge invariance. But the
resulting impact of the RCs on the measured cross sections depends
on the measurement performance i.e. reconstruction of the
kinematical variables, detector properties etc.

It is interesting to notice that there is a new proposal of the
neutrino project nuSTORM \cite{Kyberd:2012iz}. This experiment is
going to measure the $\nu_e$ and $\nu_\mu$ scattering cross
sections. It should allow to critically investigate the
systematical differences between $\nu_e$ and $\nu_\mu$ cross
sections.

To summarize, we have obtained the TBE correction, its hard photon
contribution,  to the CCQE cross sections.  The TBE effect is two
times larger for $\nu_e$ than for $\nu_\mu$. The relative
difference between the TBE correction to $\nu_e$ and $\nu_\mu$
cross sections is of the order of $2\%$.   In the low $Q^2$ limit
it increases, while for the large values of $Q^2$ it reduces the
cross sections. Eventually the other QED-like corrections have
been obtained. It turned out that  the TBE effect cancels a
non-negligible  part of the soft photon Bremsstrahlung
contribution.

\section*{Acknowledgements}
We thank Jan Sobczyk for the valuable remarks to the paper.

A part of calculations has been carried out in Wroclaw Centre for
Networking and Supercomputing 
grant No. 268.

\end{document}